\documentclass[showpacs,twocolumn,aps,preprintnumbers,letterpaper]{revtex4}
\usepackage{amsmath,amssymb}
\usepackage{epsfig}
\usepackage{graphicx}
\usepackage{amsmath}
\usepackage{amsfonts}
\usepackage{epstopdf}

\newcommand{\be}{\begin{equation}}
\newcommand{\ee}{\end{equation}}
\newcommand{\bear}{\begin{eqnarray}}
\newcommand{\eear}{\end{eqnarray}}
\newcommand{\ba}{\begin{array}}
\newcommand{\ea}{\end{array}}

\def\be{\begin{eqnarray}}
\def\ee{\end{eqnarray}}

\def\roughly#1{\mathrel{\raise.3ex\hbox{$#1$\kern-.75em%
\lower1ex\hbox{$\sim$}}}}

\begin{document}

\title{Probing Wilson Loops in AdS/QCD}

\author{Yizhuang Liu}
\email{yizhuang.liu@stonybrook.edu}
\affiliation{Department of Physics and Astronomy, Stony Brook University, Stony Brook, New York 11794-3800, USA}

\author{Ismail Zahed}
\email{ismail.zahed@stonybrook.edu}
\affiliation{Department of Physics and Astronomy, Stony Brook University, Stony Brook, New York 11794-3800, USA}

\date{\today}

\begin{abstract}
We use AdS/QCD to analyze the quark and gluon scalar and pseudo-scalar
condensates around static color sources described by a circular Wilson loop.
We also derive the static dipole-dipole interactions between rectangular Wilson loops in AdS/QCD and 
discuss their relevance for static string interactions in  QCD
at strong coupling.
\end{abstract}
\pacs{11.25.Tq, 13.60.Hb,13.85.Lg}

\maketitle

\setcounter{footnote}{0}


\section{Introduction}

QCD is the fundamental theory of strong interactions. It has proven challenging in the infrared
as its fundamental constituents are confined. In this regime, the quarks and gluons
interact strongly and form strongly interacting hadrons. Holographic QCD is  an attempt
to solve QCD at large number of colors $N_c$ and strong t$^\prime$ Hooft coupling $\lambda=g^2N_c$,
guided by the gauge/gravity duality observed in string theory~\cite{Maldacena:1998im}.

The gauge/gravity principle states that in the double limits of a large number of colors or strong coupling
($N_c\gg \lambda\gg 1$) supersymmetric gauge
 theory is equivalent to a one-higher dimensional gravity theory coupled to some bulk fields that are dual to
 gauge invariant operators of QCD. The conformal nature of the gauge theory is encoded in an AdS space
 for gravity. Each of the propagating field in bulk AdS is in one-to-one correspondence with an operator in the
 field theory. Although the correspondence is established for type IIB superstring theory in $AdS_5\times S_5$,
 it is believed to hold for string theory in a general background.

 The string theory in curved backgrounds is in general  difficult to solve. However, at strong
 coupling the string theory turns to a weakly coupled classical supergravity which is
 tractable. The best established gauge/gravity dual correspondence is for ${\cal N}=4$ super-Yang-Mills
 theory, which on the gravity side is described by a stack of $N_c$ D3 branes sourcing an $AdS_5$
 metric in bulk. In so far, there is no exact gravity dual candidate of pure Yang-Mills or QCD. The closest
 dual proposal from string theory is the Witten-Sakai-Sugimoto (WSS) model
 \cite{Witten:1998zw,Sakai:2004cn} based on a stack of D4 branes
 with probe D8 branes using the so-called top-down approach. A more phenomenological or bottom-up
 approach was originally suggested by Erlich-Katz-Son-Stephanov (EKSS) \cite{Karch:2006pv} and others.
 We will refer to it as AdS/QCD. 
 
  In this paper we will use the bottom-up approach to analyze the gluonic and fermionic condensates around
 static color sources as circular Wilson loops. We will also derive an explicit static dipole-dipole interactions between 
rectangular Wilson-loops.  The study of the spatial distribution of the quark condensates around rectangular
 Wilson loops have been carried recently on the lattice~\cite{Iritani:2014jqa} to understand the faith of the chiral pairing in the
 vicinity of a flux tube. These studies aim at  probing the interplay between the spontaneous breaking of chiral
 symmetry and confinement, two key aspects of the QCD vacuum.  Recent phenomenological analyses of 
~\cite{Iritani:2014jqa} suggest the presence of a $\sigma$-meson cloud around the QCD string
\cite{Kalaydzhyan:2014tfa}.  A model analysis in the Schwinger model was also suggested in~\cite{Kharzeev:2014xta}.

 Finally, the nature and strength of two small dipole-dipole
 interactions in QCD may shed light on the character of the static  interactions between QCD strings. These interactions
 are important  in the QCD string-black-hole duality whether in thermal equilibrium or in high-energy 
 collisions~\cite{Shuryak:2013sra}
  (and references therein). Static dipole-dipole interactions have been discussed using YM instantons~\cite{Shuryak:2003rb,
 Giordano:2009vs}   and ${\cal N}=4$ SYM  \cite{Berenstein:1998ij}.

The outline of this paper is as follows:  In section 2, we formulate the model. In section 3 we define the minimal circular
 loops and their coupling to the lowest dimensional scalar. In section 4 and 5 we analyze the scalar and pseudo-scalar
 quark condensates (form factors), the  scalar and pseudo-scalar gluonic condensates 
 (form factors) around a heavy quark described by a circular 
 Wilson loop. Some critical remarks regarding our  analysis are given
 in section 6. In section 7 we construct the static dipole-dipole potential in the confined phase and discuss its qualitative structure
 in the Coulomb phase. Our conclusions are  in section 8.

\section{The model}

The soft wall version of the EKSS model consists of a 5D (flavor) gauge theory in a slice of $AdS_5$ space-time.
Assuming the gauge/gravity correspondence, one introduces bulk fields dual to the gauge-singlet operators in QCD.
Specifically

\begin{eqnarray}
S=&&\int d^4x dz\,\sqrt-g\,e^{-\phi}\,\nonumber\\
&&\times \left[-|DX|^2-m_5^2\,|X|^2-\frac 1{2g_5^2}\left(F_L^2+F_R^2\right)\right]
\label{1}
\end{eqnarray}
with the dilaton background  $\phi\equiv k^2z^2$, $DX=dX-iA_LX+iXA_R$ and the
$AdS_5$ gravity background metric

\begin{equation}
ds^2=\frac 1{z^2}(-dt^2+dx^2+dz^2)
\label{1XX}
\end{equation}
in units where the AdS radius is set to 1. The soft scale $k$ is set by the rho meson trajectory
$m_n^2=4(n+1)k^2$  after fitting the rho mass or $k=m_\rho/2=385$ MeV for $n=1$~\cite{Grigoryan:2007my}.
Throughout, this scale will be set to 1 and restored when needed by inspection.
The gravity dual of any confining gauge theory should have a bulk geometry that caps off at at a finite distance
in the holographic direction as first suggested by Polshinski and Strassler~\cite{Polchinski:2002jw} in the so-called hard wall model. Here,
the dilaton background enforces that softly through a quadratic profile or soft wall. 

The bulk fields correspond to the following QCD operators at the boundary

\begin{eqnarray}
X_{ij} \rightarrow&& \overline{q}_{Li}q_{Rj}\nonumber\\
A_{L,R\mu}^a\rightarrow&& \overline{q}\gamma_\mu(1\pm \gamma_5)\tau^aq
\end{eqnarray}
which are the flavor scalars and left-right vector fields. A comparison of the correlators on the boundary and in bulk shows that the
bulk mass is related to the scaling dimension $\Delta$ and spin $p$ of the boundary operators. Specifically

\be
m_5^2=(\Delta-p)(\Delta+p-4)
\label{2}
\ee
For the scalar quark operator $\bar q q$  in QCD:  $\Delta =3$ and $m_5^2=-3$. For both the scalar  and pseudo-scalar
 gluon operators  $F^2,F\tilde{F}$ in QCD:  $\Delta =4 $ and $m_5^2=0$ (see below). We note that the use of a warped
 background that accounts for the breaking of conformality in modified AdS/QCD changes the field strengths anomalous 
 dimensions~\cite{Powell:2012zj}. It will not be pursued here.

\section{Heavy color source as a Circular Wilson loop}

We model a heavy color source on the boundary through a circular Wilson loop of radius $a$. In the gauge/gravity
correspondence the Wilson loop is described by a minimal Nambu-Goto (NG) string in bulk with a circle as a boundary.
For small $a$ in units of the AdS radius (set to 1 here) the minimal surface is

\begin{equation}
x=\sqrt{a^2-z^2}\,{\rm cos} \varphi
\end{equation}
\begin{equation}
y=\sqrt{a^2-z^2}\,{\rm sin} \varphi
\end{equation}
The surface area in form notation  is
\begin{equation}
dA=\frac{adzd\varphi}{z^2}
\end{equation}

The coupling of the NG string as a circular loop of size $a$ to the flavor scalar is not known in the soft wall model. Since
operators in QCD are only identified in bulk by their anomalous dimension $\Delta$ and spin $p$ we suggest that the bulk
scalar-to-Wilson-loop coupling in the soft-wall model can be borrowed from the analogous coupling in ${\cal N}=4$ SYM.
In the latter, various bulk fields follow from the KK reduction of type IIB supergravity on $AdS_5\times S_5$. In particular
the trace of the graviton (the dilaton) contributes a scalar $X$ in bulk with the same dimension and free mass $m_5^2=-3$
\cite{Berenstein:1998ij}. Since the graviton couples naturally to the string world-sheet, so does its trace
through ~\cite{Berenstein:1998ij}

\begin{equation}
\frac{1}{2\pi\alpha N_c}\int dA(-6X)\frac{z^2}{a^2}
\label{2X}
\end{equation}
with $6=2\Delta(\bar q q)$. Here $\alpha=l_s^2$ and the string tension $\sigma_T=1/2\pi\alpha$.  
The string coupling is $g_s\equiv 4\pi\lambda/N_c$ with $l_s^4\equiv \lambda$ in (walled) $AdS_5$.

\section{Quark Condensates around a Heavy Color Source}

It is now straightforward to estimate the amount of quark  condensate around a heavy color source represented by
the small circular Wilson loop described in the preceding section. Specifically, the connected quark condensate is

\begin{equation}
\frac{\left< \bar q q(x)W(C)\right>_c}{\left<W(C)\right>}= \frac{1}{2\pi\alpha N_c} \int dA\,\frac{-6z^2}{a^2z_*^3}G(x,z;x_*,z_*)
\label{4XX}
\end{equation}
where $(x_*,z_*\rightarrow 0)$ is the position of the quark operator on the boundary. (\ref{4XX}) is readily understood as the
scalar quark form factor of the Wilson loop.

The vacuum solution to the soft-wall
EKSS model (\ref{1}) describes a scalar condensate. For that we set $A_L=A_R=0$ and choose $\Delta=3$ and $p=0$
to describe the quark condensate  $\overline{q}q$, so that $m_5^2=-3$. The equation of motion is

\be
\frac{d}{dz}(e^{-B}\frac{d}{dz}X_0)+3\frac{e^{-B}}{z^2}X_0=0
\label{2XX}
\ee
subject to the ultra-violet boundary condition

\be
X_*(z)=m\,z+\left<\overline{q}q\right>\,z^3+{\cal O}(z^4)
\label{qq}
\ee
(\ref{2XX}) admits a unique solution. The actual form of $X_*(z)$  is not needed for the
analysis of (\ref{4XX}) since the fluctuations around it decouple thanks to the quadratic
nature of the action in $X$ in (\ref{1}).

 \begin{figure}[htb]
\centerline{
\includegraphics[width=4cm]{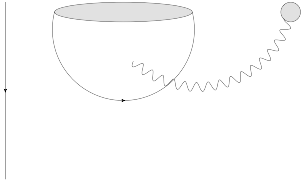}}
\caption{Circular Wilson loop probed by $X\equiv\left<\bar q q \right>$ at the boundary. See text. }
\label{DD1}
\end{figure}

With this in mind, the Green function for the scalar field $X$ in bulk admits a Fourier transform

\begin{equation}
G(x,z;x_*,z_*)=\int \frac{d^4p}{(2\pi)^4}e^{-ip\cdot (x-x_*)}\,iG(p;z,z_*)
\label{4}
\end{equation}
owing to space translational invariance, with a mode decomposition

\be
G(p; z, z_*)=\sum_{n=0}^{\infty}\frac{-iX_n(z)X_{n}(z_*)}{p^2+m^2_{Xn}}
\label{4X}
\ee

The modes in (\ref{4X}) are solutions to the equation of motion for the X-field in bulk
after substituting $X\rightarrow e^{-ip\cdot x}X$ for $p^2=m_X^2$ or

\begin{equation}
\frac{d}{dz}(e^{-B}\frac{d}{dz}X)+3\frac{e^{-B}}{z^2}X+m_X^2\,e^{-B}X=0
\end{equation}
with $B=z^2+3\,{\rm ln}z$. Following~\cite{Karch:2006pv} we define $X=e^{B/2}\,\psi$ and solve for $\psi$.
Thus

\begin{equation}
X_n=e^{B/2}\psi_n=z^3\,\sqrt{\frac{2}{1+n}}L^1_n(z^2)
\label{6}
\end{equation}
with
\begin{equation}
m^2_{Xn}=4n+\frac{9}{2}
\label{6X}
\end{equation}

Using (\ref{4X}-\ref{6}) into (\ref{4}) one can undo the p-integration (\ref{4}) in terms of Bessel functions.
For equal times or $x=(0, {\bf x})$
\be
\int\frac{d^4p}{(2\pi)^4}\frac {e^{-ip\cdot x}}{p^2+m_{Xn}^2}=\frac {m_{Xn}}{4\pi^2 |{\bf x}|} {\bf K}_1(m_{Xn} |{\bf x}|)
\ee
Thus, 
\begin{eqnarray}
&&\frac{\left< \bar q q(x) W(C)\right>_c}{\left<W(C)\right>}=-12\frac{\sqrt{\lambda}}{N_c}\sum_{n=0}^{\infty}\int_{0}^{a} k^6 z^3L_n^1(z^2k^2)\nonumber\\
&&\times\,\frac{m_{Xn}\,{\bf K}_1(m_{Xn}k\sqrt{x^2+a^2-z^2-2x\sqrt{a^2-z^2}cos\phi})}{ka\sqrt{x^2+a^2-z^2-2x\sqrt{a^2-z^2}cos\phi}}dzd\phi
\nonumber\\
\end{eqnarray}
Assuming that the distance $|x-x_*|\approx |0-{\bf x}_*|\equiv L$ between the center of the loop and the probing scalar is large,
we may ignore the details of the shape of the loop and set $L_n^1(z^2)\approx L_n^1(0)=n+1$.  The result
after the z-integration is ($L/a\gg 1$)

\begin{eqnarray}
&&\frac{\left< \bar q q(x)W(C)\right>_c}{\left<\overline{q}q\right>\left< W(C)\right>} \approx\nonumber\\
&&-\frac{6\pi k^6a^3}{\left<\overline{q}q\right>} \frac{\sqrt{\lambda}}{N_c}\sum_{n=0}^{\infty}\frac{(n+1)\sqrt{m_{Xn}}}{Lk} {\bf K}_1({m_{Xn}}kL)
\label{7}
\end{eqnarray}
where the soft wall scale $k=m_\rho/2$ was re-instated.  
The exponential fall-off caused by the confining soft wall
in (\ref{7}) is to be contrasted with the power fall-off as $1/L^6$ for ${\cal N}=4$  SYM~\cite{Berenstein:1998ij}
(see also below).
The lowest scalar mass in the exponent is $m_{X0}k=3m_\rho/2\sqrt{2}\approx 817$ MeV. Recall that in
AdS/QCD, $\left< \bar q q\right>$ is an input through (\ref{qq}).

The pseudo-scalar form factor  $\bar q i \gamma_5 q$ of the Wilson loop  in AdS/QCD involves the dual axion field 
$\xi (x,z)$ in bulk~\cite{Gursoy:2007cb}. The axion contribution to the bulk action is similar to the 
action for the $X$ field above but only $1/N_c^2$ suppressed. Therefore, we may just assume
that in AdS/QCD the bulk action for the axion field is $1/N_c^2$ that of the scalar
in (\ref{1}).  The axion coupling to the string-world sheet is also $1/N_c^2$ suppressed with
respect to (\ref{2X}) or

 \begin{equation}
 S_\xi=\frac{1}{2\pi \alpha N_c^3} \int dA (-6\,\xi)\frac{z^2}{a^2}
 \label{77XXX}
 \end{equation}
A rerun of the preceding arguments for the scalar form factor gives

\begin{eqnarray}
&&\frac{\left< \bar q i\gamma_5 q(x)W\right>}{\left< W(C)\right>} \approx \nonumber\\
&&-{6\pi k^6a^3}\frac{\sqrt{\lambda}}{N^3_c}\sum_{n=0}^{\infty}\frac{(n+1)\sqrt{m_{Xn}}}{Lk} {\bf K}_1({m_{Xn}}kL)
\nonumber\\
\label{77XX}
\end{eqnarray}
which is $1/N_c^2$ suppressed in comparison to (\ref{7}).

\section{Gluon condensates around a heavy color source}

To probe the scalar gluon condensate around a heavy quark source, we proceed through similar arguments
by replacing $\overline{q}q$ with $(gF)^2\equiv F^2$ (RGE invariant). 
In bulk, the dual of the latter is a dilaton fluctuation $\varphi$. Recall that in the
bottom-up approach, the dilaton background is an input and not a solution to the coupled gravitational equations.
This notwithstanding, the form of the action for $\varphi$  is
analogous to (\ref{1}) for $X$ but with $m_5^2=0$ as noted earlier,

\be
S_\varphi=\int d^4x dz\,\sqrt-g\,e^{-\phi}(-|D\varphi|^2)
\label{FF1}
\ee
The equation of motion for $\varphi$ proceeds as before for both the vacuum solution $\varphi_*(z)$
through  the ultraviolet-boundary condition $\varphi_*(z)=C+\left<F^2\right>z^4+{\cal O}(z^5)$, and
normal modes in bulk

\be
\varphi_n=z^4 \sqrt{\frac{2}{(1+n)(2+n)}}L^2_n(z^2)
\ee
with

\be
m^2_{\varphi n}=4n+13/2
\label{6XX}
\ee

The coupling of the gluonic scalar to the string world-sheet in the Einstein-frame, is analogous to (\ref{2X})

\be
\frac{1}{ 2\pi \alpha N_c}\int\,dA\,\frac 12 \varphi
\label{X6XX}
\ee
without the extra minus sign and z-warping. A rerun of the previous arguments yields
for the gluonic form factor or condensate of the Willson loop

\begin{eqnarray}
&&\frac{\left< F^2(x)W\right>_c}{\left< F^2\right>\left< W(C)\right>} \approx \nonumber\\
&&+\frac{\pi k^8a^4}{3\left< F^2\right>} \frac{\sqrt{\lambda}}{N_c}\sum_{n=0}^{\infty}\frac{(n+1)(n+2)\sqrt{m_{\varphi n}}}{Lk} {\bf K}_1({m_{\varphi n}}kL)\nonumber\\
\label{7X}
\end{eqnarray}
The leading exponential decay is governed by the lowest mass $m_{\varphi 0}k=\sqrt{13/8}\,m_\rho\approx 982$ MeV.
As indicated earlier,  the gluon condensate $\left<F^2\right>$ is an input in AdS/QCD.

The topological gluonic form factor $F\tilde {F}$ of the Wilson loop in AdS/QCD can be sought along the
same arguments as the pseudo-scalar form factor. The dual field $\chi(x,z)$ is represented by a
bulk action that is similar to (\ref{FF1}), but with a coupling

 \begin{equation}
 S_\chi=\frac{1}{2\pi \alpha N_c^3} \int dA (-8\,\chi)\frac{z^2}{a^2}
 \label{X7X}
 \end{equation}
with $8=2\Delta({F\bar F})$, by analogy with the axion coupling (\ref{77XXX}).
A rerun of the preceding arguments for the topological form factor of the circular Wilson loop yields

\begin{eqnarray}
&&\frac{\left< F\tilde{F}(x)W\right>}{\left< W(C)\right>} \approx \nonumber\\
&&-\frac{4\pi{k^8a^4}}{5}\frac{\sqrt{\lambda}}{N_c^3}\sum_{n=0}^{\infty}\frac{(n+1)(n+2)\sqrt{m_{\varphi n}}}{Lk} {\bf K}_1({m_{\varphi n}}kL)\nonumber\\
\label{7XX}
\end{eqnarray}
which is opposite to (\ref{7X}) and $1/N_c^2$ suppressed. The depletion of the topological charge near the 
Wilson loop with a strong flux-sheet is plausible, justifying a posteriori the negative coupling in (\ref{X7X})
by analogy with the axion coupling.

\section{Remarks}

In deriving (\ref{7}) using the soft wall model (\ref{1}) we have made some assumptions:
1) The dual $X$ of $\overline {q}q$ has no back-reaction on the space-time geometry;
2) Its coupling to the NG world-sheet is the same as in ${\cal N}=4$ SYM~\cite{Berenstein:1998ij}.
Some of these issues can be overcome by introducing the back-reaction of $X$ on the gravity sector
using a more realistic bottom-up approach to the gauge/gravity dual of QCD such as
V-QCD~\cite{Jarvinen:2011qe}.  Specifically, the action is now

\begin{eqnarray}
\begin{split}
S=(M^3N^3_c)\int[ \sqrt{-g}(R-\frac{4}{3}\frac{(\partial \lambda)^2}{{\lambda}^2}+V_g(\lambda)\\
-xV_f(\lambda,T)\sqrt{det(g_{ab}+h(\lambda,T)\partial_aT\partial_bT}]
\end{split}
\end{eqnarray}
with $R$ the Ricci curvature for the metric $g$, $\lambda$ the dilaton field and $T$ the tachyon field. $V_{g,f}$ are the gluonic
and sermonic potentials respectively with $x=N_f/N_c$ fixed at large $N_c$. The ensuing equations of motion naturally couple
$T$ and $g$. Recall that $T$ is dual to the quark condensate at the boundary

\begin{equation}
T_{ij}(x,z)\equiv m_{ij}(x)\,z+\sigma_{ij}(x)\,z^3+{\cal O}(z^4)
\label{8}
\end{equation}
To proceed we need to solve for $T$ subject to the boundary condition (\ref{8}), find the metric $g$ as a function of the mass
matrix $m_{ij}$. The knowledge of $g[m_ij]$ allows the construction of the minimal surface for the circular Wilson loop or
${\cal A}[m_ij]$. The scalar condensate in the presence of a Wilson loop follows then through

\begin{equation}
\left<\bar q q(x)W(C)\right>=\frac{\delta {\cal A}[m_{ij}]}{\delta m_{ij}(x)}
\end{equation}
after setting $m_{ij}(x)=m\delta_{ij}$. The local coupling to the string world-sheet will follow from

\begin{equation}
\delta {\cal A}= \int dA\, {\bf C}(x,z)\,\delta T(x,z)
\end{equation}
where ${\bf C}(x,z)$ is a local function on the surface. For $L/a\gg 1$ it is sufficient to take
${\bf C}(0,z)$ at the center of the world-sheet and use the tachyon-tachyon correlator
$\left<\delta T(0,z_*)\delta T(x,z^\prime)\right>$  for the Green function $G(0,z_*; x, z^\prime)$.
This procedure is numerically involved and will be reported elsewhere. Overall, we expect the
results to be qualitatively similar to the soft-wall result quoted above.

\section{Static Dipole-Dipole potential}

The dipole distribution around another dipole of size $a$ is best captured by the dipole-dipole
potential. For simplicity, consider static and equal size dipoles of spatial extension $a$ away
from each other by a distance $L$. In QCD this dipole-dipole potential was analyzed in the
random-instanton vacuum in~\cite{Shuryak:2003rb,Shuryak:2003ja} and more recently in ${\cal N}=4$ SYM in
\cite{Berenstein:1998ij}. We now address it in the context of the soft wall model.
By definition the dipole-dipole potential follows from the long time connected correlator

\begin{equation}
V(L)=-\lim_{T \to+\infty} \frac {1}{T}\,\,
 {\rm ln} \left(\frac{\left<W(L)W(0)\right>_c}{\left<W(L)\right>\left<W(0)\right>}\right)
 \label{12}
\end{equation}
with $W(0)$ and $W(L)$  two identical and rectangular Willson loops of width $a$ and infinite time-extent $T$,
centered at $0$ and $L$ respectively.

In AdS/QCD and for $L/a\gg1$ and small dipole sizes $a$, the minimal surface in the $x,z$
plane is unaffected by the soft wall. Thus its shape is given by

\begin{equation}
(\frac{dz}{dx})^2=\frac{z^4_h}{z^4}-1,|x-x_o|<a/2,z<z_h,z_h\propto a
\end{equation}
with $x_0$ the center of the rectangular loop.

The scalar coupling $X$ to the rectangular Wilson loops now reads

\be
\frac{1}{2\pi\alpha N_c}\int dA\frac{z^2}{z^2_h}(-6X)
\label{13}
\ee
instead of (\ref{2X}). For $L/a\gg1$  we can use arguments similar to those developed earlier to
re-express the connected correlation function in (\ref{12}) in terms of the scalar propagator (\ref{4})
folded with the couplings to the respective couplings (\ref{13}) on the world-sheets. The result
for the potential is

\begin{equation}
V(L)\approx  -4\pi k^5a^4 \frac{\lambda}{N_c^2}\sum_{n=1}^{\infty}(n+1)\,\frac{m_{Xn}}{kL}\, {\bf K}_1(km_{Xn}L)
\label{14}
\end{equation}
with $k=m_\rho/2$. The overall minus sign follows from (\ref{12}). The scalar exchange is attractive and exponentially suppressed in
the confined phase. The result  (\ref{14}) is to be compared to $-{a^4}/{L^5}$ in ${\cal N}=4$ SYM~\cite{Berenstein:1998ij}
(see below).

Although we only considered the exchange of the lowest dimensional scalar $X$ in deriving (\ref{14}) additional
exchanges in the form of gravitons and  B-fields are also expected. In the soft-wall model considered
here they are characterized by similar couplings but their boundary squared masses are all larger than the $9k/2$ for
the scalar in (\ref{6X}). Indeed, the squared mass of the dilaton is $13k/2$ as in (\ref{6XX}), while that of the graviton is $8k$
both of which are heavier.

\begin{figure}[htb]
\centerline{
\includegraphics[width=4cm]{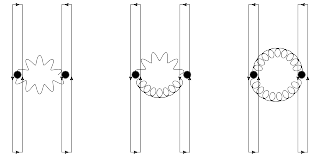}}
\caption{Static dipole-dipole interactions: $\Phi \Phi$ exchange (left), $\Phi F$ exchange (middle) and $FF$ exchange (right). See text. }
\label{DD1}
\end{figure}

It is instructive to derive the results in the
non-confining or Coulomb phase using the qualitative arguments developed in~\cite{Shuryak:2003rb,Shuryak:2003ja}. Indeed, in ${\cal N}=4$
the composite coupling for the $\Delta=2$ operator made of two adjoint scalars $\Phi\Phi$ each with coupling $g\Phi$ to a heavy quark or Wilson-loop,
follows from second order perturbation theory or ${|<0|g\Phi|1>|^2}/\Delta E \approx (a \sqrt{\lambda}/{N_c})(\Phi\Phi)$
with the energy splitting $\Delta E\approx \sqrt \lambda/{a}$ at strong coupling~\cite{Maldacena:1998im}. Inserting these operators between a heavy-quark-antiquark pair at a distance $L$
as shown in Fig.~\ref{DD1} (left) yields the potential

\be
V_{\Phi\Phi}(L)\approx -\left(\frac{a\sqrt{\lambda}}{N_c}\right)^2\int\,d\tau\, \left(\frac 1{L^2+\tau^2}\right)^2\approx -\frac {\lambda a^2}{N_c^2L^3}\nonumber\\
\label{15}
\ee
The composite coupling for $\Delta=4$ operator made of say two adjoint electric
gluons $E^2$ each with coupling $gaE$ is $a^3\sqrt{\lambda}/N_c\,E^2$. From Fig.~\ref{DD1} (right), a rerun of the preceding arguments  yield the potential

\be
V_{FF}(L)\approx- \left(\frac{a^3\sqrt{\lambda}}{N_c}\right)^2\int\,d\tau\, \left(\frac 1{L^2+\tau^2}\right)^4\approx -\frac {\lambda a^6}{N_c^2L^7}\nonumber\\
\label{16}
\ee
The composite coupling for
$\Delta=3$ follows from the cross couplings in second order perturbation theory or
$|\left<0|(g\Phi+ga\cdot E)|1\right>|^2/\Delta E$ leading to ${\sqrt{\lambda} a^2}/{N_c}\Phi E$. The dipole-dipole potential is then

\be
V_{\Phi F}(L)\approx -\left(\frac{a^2\sqrt{\lambda}}{N_c}\right)^2\int\,d\tau\, \left(\frac 1{L^2+\tau^2}\right)^3\approx -\frac {\lambda a^4}{N_c^2L^5}\nonumber\\
\label{17}
\ee
(\ref{15}-\ref{17})  summarize to

\be
V_\Delta(L)=-{\cal C} (\Delta)\,\frac{\lambda a^{2\Delta-2}}{N_c^2L^{2\Delta-1}}
\ee
in agreement with the result in~~\cite{Berenstein:1998ij}.
The overall constant ${\cal C}(\Delta)$ is beyond the qualitative nature of our arguments.
Its precise value can be found in~~\cite{Berenstein:1998ij} using a more detailed analysis.
For completeness we quote its value

\begin{eqnarray}
&&{\cal C}(\Delta)=\frac{\Gamma(1/4)^{4\Delta-4}}{32^{2\Delta+9}\pi^{3\Delta-7/2}}\nonumber\\
&&\times \frac{(\Delta-1)^2(\Delta-2)^2(\Delta-3)\Gamma(\Delta-1/2)\Gamma(\frac{\Delta-1}{4})^4}{\Gamma(\Delta) \Gamma(\frac{\Delta-1}{2})^2}\nonumber\\
\label{1XX}
\end{eqnarray}

\section{Conclusions}

To probe flux tubes in QCD is notoriously hard outside lattice simulations.
We have suggested the gauge/gravity correspondence as a simple framework
for addressing this issue. We have represented static color charges by circular
Wilson loops and shown how to probe the analogue of the chiral and gluon
condensate around these charges.

In AdS/QCD the scale is set by the rho mass $m_\rho$. The 
quark clouds are dominated by the
exchange of a light mass of order $3m_\rho/2\sqrt{2}$, while the gluonic
clouds involves a light mass of order $\sqrt{13/8}\,m_\rho$. The quark and
pseudo-scalar gluon clouds are depleted by the heavy quark source 
(negative contribution) while the scalar gluon
cloud is enhanced by the heavy quark source (positive contribution). Their
strong coupling to the heavy
quark world-sheet as a loop of radius $a$,
is generically $(-am_\rho^2)^\Delta\sqrt{\lambda}/N_c$ with $\Delta=3,4$ for the
scalar quark and gluon insertions respectively. The pseudo-scalar gluon coupling
is $1/N_c^2$ the scalar gluon coupling.

The depletion of the scalar quark condensate 
around a  rectangular Wilson loop was noted in the recent lattice simulations~\cite{Iritani:2014jqa}.
It would be interesting to evaluate the pseudo-scalar quark, scalar and pseudo-scalar gluon condensates in the
same simulations for comparison with our results.

The interaction between small size and static dipoles in the present holographic construction,
provides us with some insights to the nature of the static interactions between QCD strings
in both the confined and Coulomb phase. In AdS/QCD, the interaction is attractive
and dipole-like in the Coulomb phase, of the form $-\lambda\, a^{2\Delta-2}/N_c^2L^{2\Delta-1}$.
 In QCD, the dominance of the $\Delta=3$
or quark exchange ($\overline{q}q$) and $\Delta=4$ or glueball exchange ($F^2$) are likely
to saturate the dipole-dipole exchange at large distances with an edge for  the light
scalar exchange or $\Delta=3$~\cite{Kalaydzhyan:2014tfa}. Dynamical strings, involve
both static and non-static or velocity-dependent potentials. The latter are outside the scope of this work.

\section{Acknowledgements}

We would like to thank Ionannis Iatrakis and Edward Shuryak for discussions.
This work was supported by the U.S. Department of Energy under Contracts No.
DE-FG-88ER40388.

 \vfil

\end{document}